\documentclass[aps,twocolumn,superscriptaddress]{revtex4}
\usepackage{epsfig}
\usepackage{color}
\usepackage{graphicx}
\usepackage{dcolumn}
\usepackage{bm}
\usepackage[normalem]{ulem}
\usepackage{textcomp}
\usepackage{amssymb}

\begin{document}

\title{Orbital-selective coherence-incoherence crossover and metal-insulator transition in Cu-doped NaFeAs}

\author{S. L. Skornyakov}
\affiliation{M. N. Miheev Institute of Metal Physics of Ural Branch of Russian Academy of Sciences, 
18 S. Kovalevskaya Street, 620137 Yekaterinburg, Russia}
\affiliation{Ural Federal University, 620002 Yekaterinburg, Russia}

\author{V. I. Anisimov}
\affiliation{M. N. Miheev Institute of Metal Physics of Ural Branch of Russian Academy of Sciences, 
18 S. Kovalevskaya Street, 620137 Yekaterinburg, Russia}
\affiliation{Ural Federal University, 620002 Yekaterinburg, Russia}

\author{I. Leonov}
\affiliation{M. N. Miheev Institute of Metal Physics of Ural Branch of Russian Academy of Sciences, 
18 S. Kovalevskaya Street, 620137 Yekaterinburg, Russia}
\affiliation{Ural Federal University, 620002 Yekaterinburg, Russia}

\date{\today}

\begin{abstract}
We study the effects of electron-electron interactions and hole doping on the 
electronic structure of Cu-doped NaFeAs using the density functional theory 
plus dynamical mean-field theory (DFT+DMFT) method. In particular, we employ 
an effective multi-orbital Hubbard model with a realistic bandstructure of 
NaFeAs in which Cu-doping was modeled within a rigid band approximation and 
compute the evolution  of the spectral  properties, orbital-dependent electronic 
mass renormalizations, and magnetic properties of NaFeAs upon doping with Cu. 
In addition, we perform fully charge self-consistent DFT+DMFT calculations 
for the long-range antiferromagnetically ordered Na(Fe,Cu)As with Cu $x=0.5$ 
with a real-space ordering of Fe and Cu ions. Our results reveal a crucial 
importance of strong electron-electron correlations and local potential 
difference between the Cu and Fe ions for understanding the \textbf{k}-resolved 
spectra of Na(Fe,Cu)As. Upon Cu-doping, we observe a strong orbital-dependent 
localization of the Fe $3d$ states accompanied by a large renormalization of 
the Fe $xy$ and $xz$/$yz$ orbitals. Na(Fe,Cu)As exhibits bad metal behavior 
associated with a coherence-to-incoherence crossover of the Fe $3d$ electronic 
states and local moments formation near a Mott metal-insulator transition (MIT). 
For heavily doped NaFeAs with Cu $x \sim 0.5$ we obtain a Mott insulator with a 
band gap of $\sim$0.3 eV characterized by divergence of the quasiparticle 
effective mass of the Fe $xy$ states. In contrast to this, the quasiparticle 
weights of the Fe $xz$/$yz$ and $e_g$ states remain finite at the MIT. The MIT 
occurs via an orbital-selective Mott phase to appear at Cu $x\simeq0.375$ with 
the Fe $xy$ states being Mott localized. We propose the possible importance of 
Fe/Cu disorder to explain the magnetic properties of Cu-doped NaFeAs.
\end{abstract}

\maketitle

\section{INTRODUCTION}
The discovery of unconventional superconductivity in high-$T_c$ cuprates and 
in Fe-based pnictides and chalcogenides (Fe-based superconductors, FeSCs) has 
received enormous attention over the past several decades \cite{review_cusc,review_fesc}. 
The high-$T_c$ cuprates and FeSCs show many similarities, e.g., in the vast 
class of FeSCs and cuprates superconductivity appears as a result of the 
suppression of a long-range, antiferromagnetic (AFM) or nematic 
phase~\cite{review_cusc,review_fesc,nematicity}. It has been proposed that 
antiferromagnetic spin fluctuations play a decisive role in the mechanism of 
the $s^\pm$ (in FeSCs) and $d$-wave (in cuprates) high-$T_c$ superconductivity \cite{review_cusc,review_fesc,s_pm,s_pm_2,nematicity,pairing_cusc,Sprau_Science_2017}. 
In addition, both FeSCs and cuprates reveal the crucial importance of strong 
correlations which favor to electronic localization and severe (orbital-selective) 
quasiparticle mass renormalizations~\cite{FeSCS_correlations,Weber_NatPhys_2010,PhysRevLett.110.146402,dftdmft_fech,Deng_PRL_2013,Medici_PRL_2014,Yu_PRB_2011}.

At the same time, the electronic properties of the parent phases of FeSCs and 
cuprate superconductors are significantly different. In fact, the parent phase 
of cuprates is a Mott (or charge transfer) insulator with localized magnetics 
moments~\cite{review_cusc,Imada_review}. Its doping leads to a Mott insulator-metal 
transition which is followed by the emergence of high-$T_c$ superconductivity. 
In contrast to that FeSCs are bad metals in their parent phase~\cite{review_fesc,FeSCS_correlations,PhysRevLett.110.146402,Medici_PRL_2014,Yu_PRB_2011,Deng_PRL_2013}. 
This suggests an intermediate range of correlation strength in FeSCs and seems 
to point out itinerant nature of their magnetic moments. The latter has been 
attributed to the multiorbital character of the electronic structure of FeSCs, 
in stark contrast to the one-band behavior of cuprates. Although bad-metal 
behavior and large band renormalizations in FeSCs can in principle be explained 
by the proximity to an orbital-selective Mott phase, no correlated insulating 
state has been reported in FeSCs despite significant research 
efforts~\cite{Yu_PRB_2011,Nica_npj_2017,Medici_PRL_2014}. For quite a long time 
it was unclear whether or not a correlated (Mott or charge transfer) insulator 
can be realized in the phase diagram of FeSCs.

In order to address this question Song \emph{et al.} conducted a detailed 
experimental study of the iron pnictide NaFeAs doped with Cu~\cite{Song_ncomm_2016}. 
It was established that the heavily-doped Na(Fe$_{1-x}$Cu$_x$)As with $x\sim0.5$ 
exhibits a real space ordering of Fe and Cu ions and makes a phase transition in 
a Mott insulating state. Na(Fe$_{1-x}$Cu$_x$)As with $x\sim0.5$ exhibits insulating 
behavior in the dc resistivity up to room temperature with an activation energy 
of $\sim$100 meV~\cite{Song_ncomm_2016,Matt_PRL_2016,Wang_PRB_2013}. Below the 
N\'eel temperature $T_{\rm N}\simeq 200$~K the insulating phase concurs with a 
long-range AFM $(1,0,\frac{1}{2})$ ordering. The insulating behavior was found 
to persist well above $T_{\rm N}$, implying that Na(Fe,Cu)As with Cu $x\simeq0.5$ 
is a Mott insulator, in accordance with scanning tunneling microscopy measurements 
near $x=0.3$~\cite{Cun_PRX_2015}. Moreover, consistent with a more local origin 
of magnetism in Na(Fe,Cu)As with $x\simeq0.5$, the ordered magnetic moment of Fe 
ions in AFM NaFe$_{0.5}$Cu$_{0.5}$As $\sim$1.1 $\mu_{\rm B}$ is sufficiently 
higher than that in the spin-density wave AFM phase of NaFeAs, $\sim$0.09$-$0.32 
$\mu_{\rm B}$~\cite{Wright_PRB_2012,Park_PRB_2012,Li_PRB_2009}. It is also 
interesting to note that the N\'eel temperature in NaFeAs (Cu $x=0$) is relatively 
small, $T_{\rm N}\sim 40$~K~\cite{Park_PRB_2012,Li_PRB_2009,Zhang_PRL_2014}. In fact, 
it is significantly smaller than that in the heavily Cu-doped NaFeAs, implying 
the rise of magnetic correlations in NaFeAs upon Cu doping.

Most notably, upon decreasing $x$ from 0.5, the value of the ordered magnetic 
moment was found to gradually decrease until bulk superconductivity emerges 
below $x\sim0.05$, with a critical temperature $T_c \simeq 11$~K~\cite{Wang_PRB_2013,Tan_PRB_2017}. 
In this respect Na(Fe$_{1-x}$Cu$_x$)As is a unique system among FeSCs in which 
superconductivity seems to be ``smoothly'' connected to a Mott-insulating state, 
implying the importance of electron correlations for sustaining of the high-$T_c$ 
superconductivity in FeSCs. In addition, angle-resolved photoemission (ARPES) 
measurements combined with the electronic band structure calculations within 
the DFT+$U$ method have shown that at $x\sim 0.44$ Na(Fe$_{1-x}$Cu$_x$)As 
is a narrow-gap insulator with the energy gap originating from the on-site 
Coulomb interactions of the Fe $3d$ orbitals~\cite{DFTpU,Matt_PRL_2016}. This 
behavior has been confirmed by the DFT+dynamical mean-field theory (DFT+DMFT) 
analysis given by Charnukha~\emph{et al.}~\cite{Charnukha_PRB_2017,dftdmft_nsc}. 
In particular, it was shown that mutual agreement between the theoretical and 
experimental ARPES spectra can be significantly improved by taking into account 
the dynamical on-site Coulomb correlations within DFT+DMFT. Based on a detailed 
comparison of optical spectroscopy and DFT+DMFT results, the authors proposed 
that Na(Fe,Cu)As is a correlated Slater insulator, characterized by the crossover 
from a correlated-insulator to metal phase with highly incoherent charge transport 
due to large fluctuating moments. Moreover, nuclear magnetic resonance measurements 
of the magnetic phase of Na(Fe,Cu)As for $x\leqslant 0.5$ performed by Xin~\emph{et al.} 
reveal the possible existence of defects of the Fe and Cu stripes in 
Na(Fe,Cu)As~\cite{Xin_PRB_2020}. This result suggests that the electronic state 
of Na(Fe$_{1-x}$Cu$_x$)As can also be affected by Cu/Fe disorder which plays as 
an extra mechanism promoting the correlated insulating state at $x \sim 0.5$ due 
to Anderson localization. Overall, these results demonstrate that electronic 
correlation effects in the Fe $3d$ states are an essential ingredient for understanding 
the electronic structure of Na(Fe,Cu)As.

Applications of DFT+DMFT have proven to give a good quantitative description of 
the electronic structure and magnetic properties of various FeSCs, including 
NaFeAs~\cite{dmft_fesc, U_in_superconductors,dftdmft_fech}. However, these 
investigations mostly deal with the electronic structure and magnetic properties 
of FeSCs in their normal metallic state, while the studies of a Mott insulating 
phase in the phase diagram of FeSCs are still open to debate. In our work, we 
explore the effects of electron-electron interactions and hole doping (substitution 
of Fe with Cu) on the electronic structure and magnetic properties of Cu-doped 
NaFeAs. We employ an effective multi-orbital Hubbard model with a realistic 
bandstructure of NaFeAs in which Cu-doping was modeled using a rigid band approximation. 
We use DMFT to compute the evolution of the spectral properties, orbital-dependent 
quasiparticle band renormalizations $m^*/m$, local spin susceptibilities, and 
symmetry of spin fluctuations of NaFeAs upon doping with Cu. In addition, we 
perform DFT+DMFT calculations for the long-range antiferromagnetically ordered 
Na(Fe,Cu)As with Cu $x=0.5$. In this calculation we consider NaFe$_{0.5}$Cu$_{0.5}$As 
supercell with real-space ordering of Fe and Cu ions as determined from x-ray 
diffraction~\cite{Song_ncomm_2016}. Our results reveal a crucial importance of 
strong electron-electron correlations and local potential difference between 
the Cu and Fe ions in Na(Fe,Cu)As. Upon Cu-doping, we observe a strong orbital-selective 
localization of the Fe $3d$ states accompanied by a large renormalization of the 
Fe $xy$ and $xz/yz$ orbitals. Na(Fe,Cu)As shows bad metal behavior associated 
with a coherence-to-incoherence crossover of the Fe $3d$ electronic states and 
local moments formation. For Cu $x>0.375$ it undergoes a Mott-Hubbard metal-insulator 
transition. 
It is found to occur via an intermediate orbital-selective Mott phase to appear 
at Cu $x \simeq 0.375$, in which the Fe $3d$ $xy$ orbital is Mott localized while 
other Fe $3d$ orbitals are metallic~\cite{PhysRevLett.110.146402}.
Moreover, our results suggest the 
possible importance of Fe/Cu disorder to explain the magnetic properties of 
Cu-doped NaFeAs. 

\section{RESULTS AND DISCUSSION}
\subsection{Model approach to PM Na(Fe,Cu)As}
\label{sec_tb}

We start our theoretical analysis of the effects of electron correlations and 
Cu doping on the electronic structure of paramagnetic (PM) Na(Fe,Cu)As by 
constructing a multi-orbital Hubbard model for stoichiometric NaFeAs (Cu $x=0$). 
For this purpose, we built up a model tight-binding Hamiltonian which explicitly 
includes the Fe $3d$ and As $4p$ valence states employing atomic-centered Wannier 
functions constructed within the energy window spanned by the Fe $3d$ and As $4p$ 
valence states  of NaFeAs~\cite{WannierH}. 
For the Fe $3d$ states the tight-binding Hamiltonian is supplemented by the on-site 
Coulomb interaction $U=3.5$ eV and Hund’s exchange coupling $J=0.85$ eV. These 
values are typical for FeSCs according to different estimations~\cite{U_in_superconductors}. 
In our calculations we employ the DFT+DMFT method~\cite{dftdmft_sc,Leonov_prb_2015} 
implemented within the plane-wave pseudopotential formalism with a gradient-corrected 
approximation in DFT~\cite{GGA}. 

\begin{figure}[t]
\centering
\includegraphics[width=0.88\textwidth,clip=true,angle=-90]{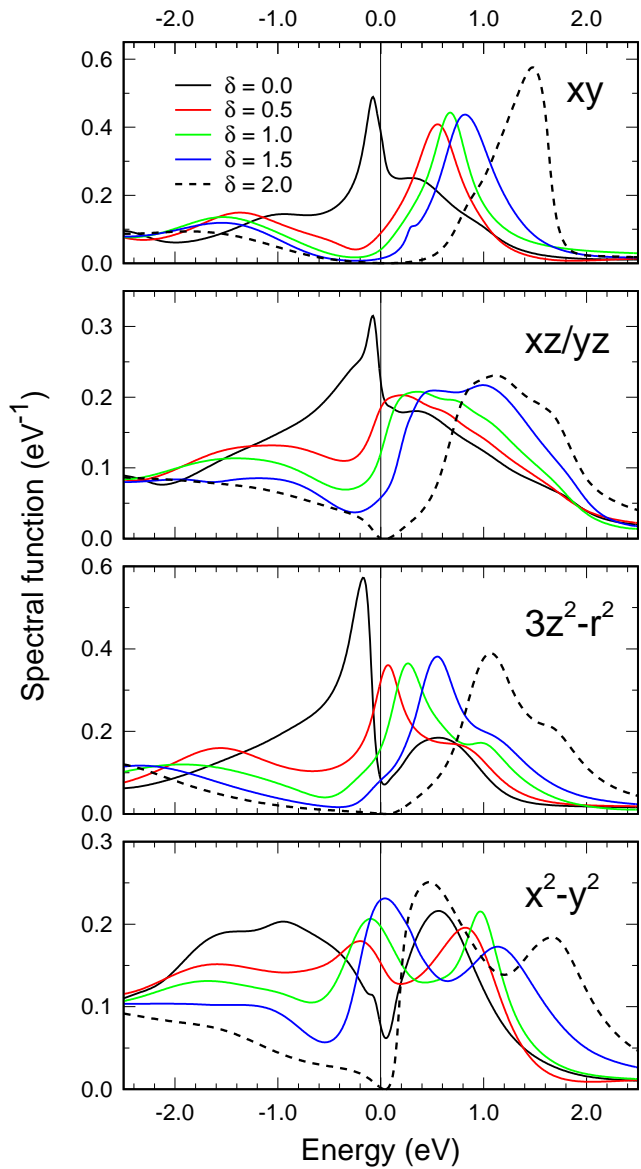}
\caption{
Orbitally-resolved Fe-$3d$ spectral functions of a tigh-binding 
model of PM NaFeAs for various hole doping $\delta$ obtained 
within DMFT at $T=290$~K.
}
\label{Fig_1}
\end{figure}

For Cu-doping $x=0$ the calculated within DFT the Wannier Fe $3d$ electron density 
is about 7.35 (per Fe ion). To model the effects of Cu doping on the electronic 
structure of Na(Fe,Cu)As we apply a rigid-band shift of the Fermi level within DFT. 
We note that in such an approach the effects of a local potential difference between 
Cu and Fe are not taken into account. We consider them explicitly in the supercell 
DFT+DMFT calculations for Cu $x=0.5$, see Sec.~\ref{sec_dftdmft}. In fact, Cu $x=0.5$ 
corresponds to the hole doping by two electrons ($\delta \equiv 2.0$) of the unit 
cell containing two formula units of NaFeAs. The DMFT many-body problem was solved 
using the hybridization expansion continuous-time (segment) quantum Monte-Carlo 
method~\cite{ctqmc}. The Coulomb interaction was treated in the density-density form 
neglecting the effects of spin-orbit coupling. We use the fully localized double-counting 
correction, evaluated from the self-consistently determined local occupations, to 
account for the interactions already described by DFT. The angle resolved spectra 
were evaluated from analytic continuation of the self-energy results using Pad\'e 
approximants~\cite{Pade}.

\begin{figure}[t]
\centering
\includegraphics[width=0.44\textwidth,clip=true,angle=-90]{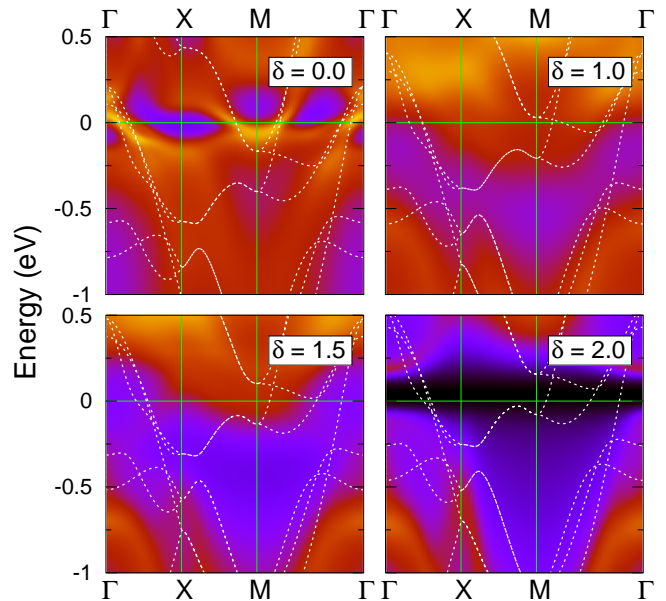}
\caption{
Electronic structure of a tight-binding model for PM Na(Fe,Cu)As along the 
$\Gamma$-X-M-$\Gamma$ path in the Brillouin zone for different 
hole doping $\delta$ calculated by DFT+DMFT at $T=290$~K (contours) 
and DFT (dashed curves).
}
\label{Fig_2}
\end{figure}

We begin with an evaluation of the electronic structure of PM Na(Fe,Cu)As. 
In Fig.~\ref{Fig_1} we display our results for the Fe $3d$ spectral functions 
computed by DMFT for the model Hamiltonian of NaFeAs upon different hole doping 
from $\delta=0$ to 2.0. Our results for the {\bf k}-resolved spectral functions 
calculated within DMFT along the $\Gamma$-X-M-$\Gamma$ path in the Brillouin 
zone (BZ) are shown in Fig.~\ref{Fig_2}. In agreement with previous results, 
for $x=0$ DMFT yields a correlated metal with the electronic structure being 
typical for FeSCs. In particular, the Fe $3d$ states are $\sim 4$~eV wide and 
show a sharp peak below the Fermi level due to the Van Hove singularity of the 
Fe $xy$ and $xz/yz$ orbitals at the BZ M-point. Our results for the Fermi surface 
are similar to those in FeSCs, with two elliptic electron pockets near the M point 
(due to the Fe $xz/yz$ and $xy$ bands) and two nearly degenerate circular hole 
pockets at the $\Gamma$ point. We note that for $x=0$ our DMFT results for the 
Fermi surface are qualitatively similar to those obtained by DFT. In addition, 
we observe a remarkable orbital-selective renormalization of the Fe $3d$ bands, 
resulting in a sizable shift (in comparison to the DFT result) of the Van Hove 
singularity of the Fe $t_2$ ($xy$ and $xz$/$yz$) bands at the BZ M-point towards 
the Fermi level. In fact, our analysis of the orbitally-resolved quasiparticle 
mass enhancement evaluated as 
$m^*/m=1 - \partial \mathrm{Im}\Sigma(\omega)/\partial\omega|_{\omega=0}$ using 
Pad\'e extrapolation of the self-energy $\Sigma(\omega)$ to $\omega\rightarrow0$ 
on the imaginary axis yields $m^*/m \sim 4.3$ and 3.5 for the Fe $xy$ and $xz/yz$ 
orbitals (see Table~\ref{Table1}). The effective mass of the Fe $x^2-y^2$ and 
$3z^2-r^2$ orbitals reveals a weaker renormalization of $\sim 2.2$--$2.6$.

\begin{table}[b]
\caption{
Orbitally resolved quasiparticle band mass enhancement $m^*/m$ in 
the tight-binding model of PM NaFeAs computed by DMFT at different 
hole doping and $T=290$~K.
}
\begin{tabular}{c|cccc}
\hline
\hline
hole doping   & $3z^2-r^2$ & $xz,yz$ & $xy$   & $x^2-y^2$ \\
\hline
$0.0$         & $2.60$     & $3.49$  & $4.34$ & $2.16$    \\
$0.5$         & $3.31$     & $3.06$  & $5.68$ & $2.03$    \\
$1.0$         & $3.43$     & $4.95$  & $8.29$ & $2.27$    \\
$1.5$         & $3.46$     & $6.14$  & $9.60$ & $2.83$    \\
\hline
\hline
\end{tabular}
\label{Table1}
\end{table}

Upon hole doping our nonmagnetic DFT results show a smooth shift of the Fermi 
level of NaFeAs. Thus, for Cu $x=0.5$ it shifts by $\sim$330 meV (see Fig.~\ref{Fig_2}). 
For Cu $x=0.5$ DFT gives a metal, in contrast to a Mott insulating behavior determined 
in the experiments (as expected due to the neglect of the effect of correlations). 
This suggests the crucial importance of electronic correlations of the Fe $3d$ 
states in NaFeAs. In agreement with this, using DMFT we obtain a large spectral 
weight transfer from low to high energies upon Cu-doping, which is accompanied 
by a metal-to-insulator phase transition for $\delta> 1.5$ (Cu $x>0.375$). It is 
accompanied by a remarkable shift of the Fe $3d$ spectral function peaks across 
the Fermi level $E_{\rm F}$ (see Fig.~\ref{Fig_1}). In particular, the peaks due 
to the $xy$ and $xz/yz$ orbitals shift above the Fermi level for $\delta>0.5$, 
while for the $x^2-y^2$ it is at about 1.5. In the same time, the spectral function 
of the $x^2-y^2$ orbital reveals different behavior upon Cu-doping. Unlike the Fe 
$t_2$ and $3z^2-r^2$ orbitals, it shows two peaks below and above $E_{\rm F}$, 
shifting to the higher energies (in the unoccupied part). 

Most notably, for $\delta> 1.5$ we observe a sharp reconstruction of the electronic 
structure of PM Na(Fe,Cu)As, associated with a Mott metal-insulator transition 
(MIT), in agreement with experiment~\cite{Song_ncomm_2016,Matt_PRL_2016}. Our results 
revel a remarkable importance of orbital selectivity in Na(Fe,Cu)As. Thus, for 
$\delta=1.5$ the most heavily renormalized Fe $xy$ orbital is seen to be insulating 
(Mott localized), whereas other Fe $3d$ orbitals are still metallic (itinerant), 
which is indicative of an orbital-selective Mott phase~\cite{PhysRevLett.110.146402}.
We obtain that Na(Fe,Cu)As with $\delta = 2.0$ ($x=0.5 $) is a Mott insulator 
with a $d$-$d$ energy gap of $0.2$~eV. We note that this result agrees well with 
the previous model DMFT calculations based on an entirely different, slave-spin, 
approach which also found a Mott insulator in the absence of a long-range 
antiferromagnetic order in heavily Cu-doped NaFeAs~\cite{Yu_PRB_2017}. At the 
same time, we observe a sizable difference in the {\bf k}-resolved spectral 
function of Na(Fe,Cu)As as compared to the ARPES measurements~\cite{Matt_PRL_2016}. 
It is presumably due to the absence of the effects of a local potential difference 
between the Cu and Fe ions in our model DMFT calculations. We note that using 
different Hubbard $U$ and Hund’s coupling $J$ does not improve the {\bf k}-resolved 
spectral functions of Na(Fe,Cu)As.

The Mott transition is accompanied by strong orbital-selective localization of 
the Fe $3d$ electrons~\cite{PhysRevLett.110.146402}. In fact, we obtain a large 
orbital-dependent enhancement of the effective mass of the Fe $3d$ states upon 
doping as shown in Table~\ref{Table1}. In particular, at the verge of a Mott 
transition, at $\delta=1.5$, $m^*/m$ is about 9.6 and 6.1 for the Fe $xy$ and 
$xz/yz$ bands, respectively. For the $e_g$ states the mass renormalizations are 
significantly weaker, $m^*/m \sim 2.8$ and 3.5 for the $x^2-y^2$ and $3z^2-r^2$ 
orbitals, respectively. This result points out that the planar $xy$ orbital is 
most renormalized, consistent with the appearance of an orbitally-selective Mott 
state \cite{Medici_PRL_2014,Sprau_Science_2017}. Moreover, our results suggest 
that the quasiparticle effective mass $m^*/m$ of the Fe $xy$ states diverges 
(i.e., Im$\Sigma(\omega)$ diverges for $\omega \rightarrow 0$) at the metal-insulator 
transition in Na(Fe,Cu)As~\cite{brinkman_rice,Leonov_prb_2015}. In contrast to 
this the $xz$/$yz$ and $e_g$ quasiparticle weights remains finite at the MIT. 
This implies the crucial importance of strong orbital-selective
correlations of the Fe $3d$ states to determine the electronic and magnetic 
properties of heavily Cu-doped NaFeAs.

\begin{figure}[t]
\centering
\includegraphics[width=0.60\textwidth,clip=true,angle=-90]{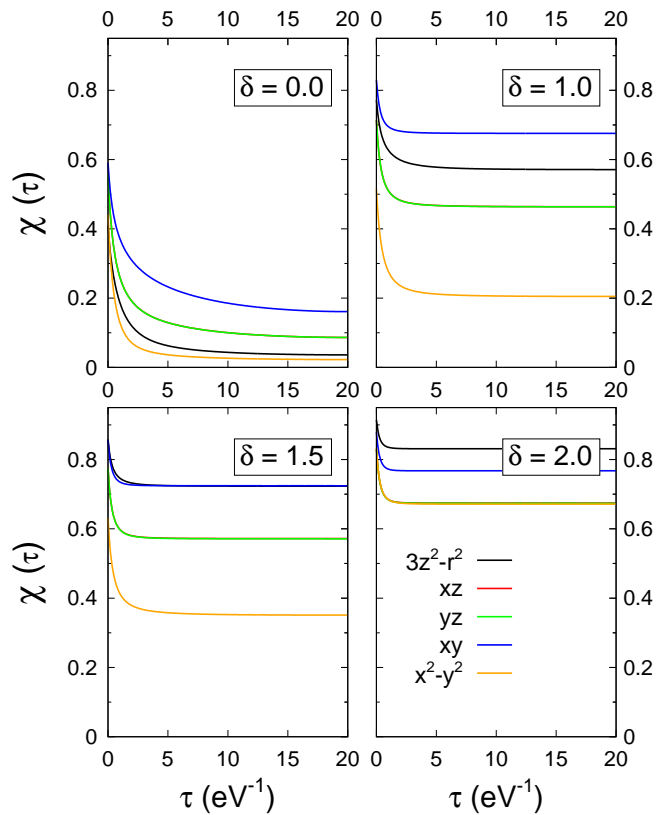}
\caption{
Orbitally-resolved local spin correlation functions
$\chi(\tau)=\left< m_{z}(\tau)m_{z}(0)\right>$ of a tight-binding 
model computed at different hole doping $\delta$ by DFT+DMFT 
at $T=290$~K.
}
\label{Fig_3}
\end{figure}

Upon hole doping, we observe a significant enhancement of incoherence of the spectral 
weight of the Fe $3d$ states, suggesting a bad metallic behavior of Na(Fe,Cu)As 
associated with the proximity to a Mott transition \cite{Imada_review,FeSCS_correlations,PhysRevLett.110.146402}. 
This behavior is accompanied by a doping-induced local moments formation in Na(Fe,Cu)As 
which results in a significant growth of the fluctuating local magnetic moments. 
In fact, upon doping from $\delta=0$ to $2.0$ the local magnetic moments increase 
from 2.3~$\mu_{\rm B}$ to 4.4~$\mu_{\rm B}$ (the corresponding fluctuating moments 
are 1.5~$\mu_{\rm B}$ and 4.3~$\mu_{\rm B}$, respectively). We therefore conclude 
that the transition is accompanied by a crossover from itinerant to localized moment 
behavior of the Fe $3d$ states. The latter is seen from our results for the 
orbitally-resolved spin susceptibility 
$\chi(\tau)=\langle \hat{m}_z(\tau) \hat{m}_z(0) \rangle$ (here, $\tau$ is the imaginary 
time) computed by DMFT for different hole doping (see Fig.~\ref{Fig_3}). Indeed, 
upon doping the Fe $3d$ electrons tend to localize to form fluctuating moments: 
$\chi(\tau)$ is seen to be almost constant, independent of $\tau$ and close to its 
maximal value. Our results also reveal strong orbital dependence of the local 
moments upon hole doping, with the $xy$ and $3z^2-r^2$ orbitals being most localized 
at high doping level. 

\begin{figure}[t]
\centering
\includegraphics[width=0.60\textwidth,clip=true,angle=-90]{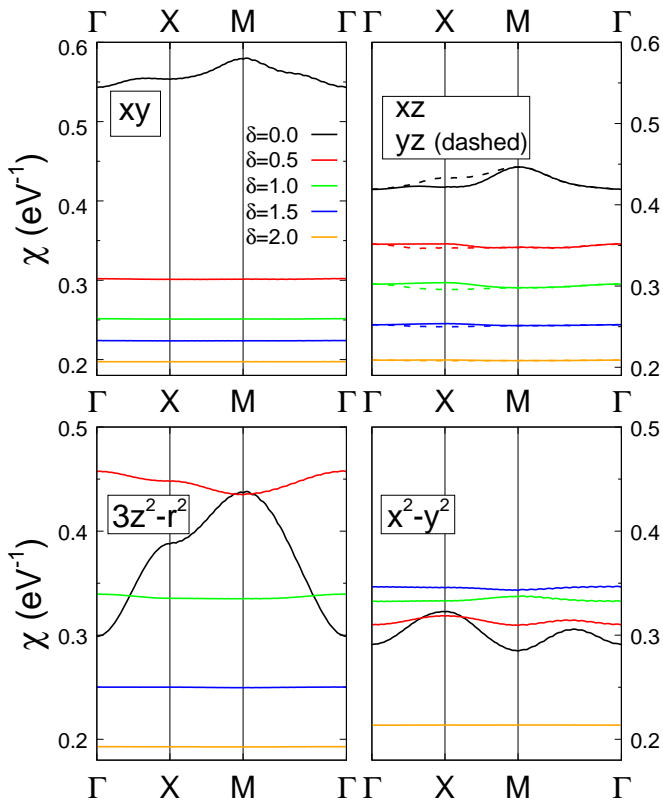}
\caption{
Orbitally resolved static spin susceptibility $\chi({\bf q})$ 
of a tight-binding model along $\Gamma$-X-M-$\Gamma$ the path as a 
function of hole doping $\delta$ computed by DFT+DMFT at $T=290$~K.
}
\label{Fig_4}
\end{figure}

It is interesting to note that upon doping (in the metallic phase) we observe 
a remarkable reconstruction of the electronic band structure of Na(Fe,Cu)As, 
associated with a change of the Fermi surface topology. It is accompanied by 
a reconstruction of magnetic correlations which can be approximately estimated 
by using the momentum-dependent static spin susceptibility $\chi({\bf q})$. 
Our result for $\chi({\bf q})$ at $x=0$ evaluated using the particle-hole bubble 
approximation shows a maximum at the BZ M-point (see Fig.~\ref{Fig_4}), which 
is characterized by an in-plane nesting wave vector $(\pi,\pi)$, consistent with 
$s^\pm$ pairing symmetry in FeSCs~\cite{s_pm,s_pm_2,Sprau_Science_2017}. This 
confirms that the leading magnetic instability of pure NaFeAs at ambient pressure 
occurs at the wave vector $(\pi,\pi)$, consistent with the spin excitation 
spectra of FeSCs~\cite{fesc_pipi}. Upon doping we observe a smooth decrease of 
$\chi({\bf q})$ for the $t_2$ orbitals which becomes almost flat and featureless 
already at $\delta=0.5$. In contrast, the shape of the $e$ orbitals susceptibility 
shows a less trivial doping dependence: at $\delta=0.5$ $\chi({\bf q})$ for the 
$3z^2-r^2$ states exhibits a sharp damping of the peak at the $M$ point which 
is accompanied by a slight increase of ferromagnetic fluctuations. $\chi({\bf q})$ 
for the $x^2-y^2$ states displays a flattening and a uniform increase followed 
by a sharp drop on the verge of the Mott transition. 

Overall, our results imply strong localization of the $3d$ electrons upon a 
doping-induced Mott metal-insulator transition in Na(Fe,Cu)As. Upon Cu-doping 
from $x=0$ to 0.5, Na(Fe,Cu)As shows a remarkable reconstruction of the 
electronic structure and coherence-to-incoherence crossover of the Fe $3d$ 
electronic states, associated with a Mott transition and the effect of local 
moments formation. This implies the crucial importance of strong correlations 
which favor to electronic localization and strong orbital-selective quasiparticle 
mass enhancement in Na(Fe,Cu)As near the Mott insulating 
phase~\cite{FeSCS_correlations,PhysRevLett.110.146402,Medici_PRL_2014,Yu_PRB_2011}. 
While the above DMFT calculations do not consider the effects of a local potential 
difference between the Cu and Fe ions, these results demonstrate that electron 
correlations in the Fe $3d$ states are an essential ingredient for understanding 
the electronic properties of Na(Fe,Cu)As. 

\subsection{DFT+DMFT calculations of AFM NaFe$_{0.5}$Cu$_{0.5}$As}

\label{sec_dftdmft}

\begin{figure}[t]
\centering
\includegraphics[width=0.27\textwidth,clip=true,angle=-90]{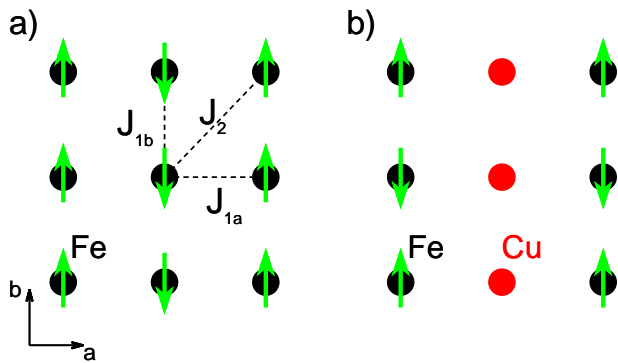}
\caption{
In-plane static magnetic configurations of (a) stoichiometric NaFeAs and 
(b) Cu-doped NaFe$_{0.5}$Cu$_{0.5}$As used in DFT+DMFT calculations. 
The dashed lines show dominating exchange paths.
}
\label{Fig_5}
\end{figure}

Next, we perform a realistic DFT+DMFT study of the electronic structure 
and magnetic properties of Na(Fe$_{1-x}$Cu$_x$)As with the real-space 
stripe-type ordering of the Fe and Cu ions as found experimentally near 
$x = 0.5$ (shown in Fig.~\ref{Fig_5}). In our calculations, we 
employ the state-of-the-art fully self-consistent in charge density 
DFT+DMFT method implemented within the plane-wave pseudopotential 
formalism~\cite{dftdmft_sc,Leonov_prb_2015}. In our DFT+DMFT calculations 
we explicitly include the Fe~$3d$, As~$4p$ and Cu~$3d$ states for the Cu 
$x=0.5$ doped Na(Fe,Cu)As by constructing a basis set of atomic-centered 
Wannier functions within the energy window spanned by these bands \cite{WannierH}. 
This allows us to take into account a charge transfer between the partially 
occupied $3d$ and $4p$ states, accompanied by the strong on-site Coulomb 
correlations of the Fe~$3d$ electrons. The Cu $3d$ states are nearly fully 
occupied with a Cu$^{1+}$ $3d^{10}$ configuration and therefore in our 
DFT+DMFT calculations we do not consider subtle correlations effects in 
the Cu $3d$ states. We use the same Hubbard $U=3.5$~eV and Hund’s rule 
coupling $J=0.85$~eV for the Fe $3d$ states as those in the model calculation 
(see Sec.~II~A). In DFT+DMFT the quantum impurity problem was solved using 
the continuous time quantum Monte Carlo (segment) method \cite{ctqmc}. 
The fully localized double-counting correction, evaluated from the 
self-consistently determined local occupations was employed. The DFT+DMFT 
calculations are performed for an antiferromagnetically ordered state of 
Na(Fe,Cu)As at a temperature $T\sim 290$~K. We use the experimentally 
established stripe configuration of the in-plane Fe moments as shown in 
Fig.~\ref{Fig_5} \cite{Song_ncomm_2016}. Note that the AFM structure 
of the Cu-doped compound is obtained from that of NaFeAs by replacing the 
ferromagnetic stripes by Cu ions. In our spin-polarized DFT+DMFT calculations 
we employ the spin-polarized DFT. Moreover, we explore the effect of Cu-doping 
on the magnetic properties of Na(Fe,Cu)As by computing the exchange 
couplings of the Heisenberg model within spin-polarized DFT+DMFT using the 
magnetic force theorem~\cite{leip}.

\begin{figure}[t]
\centering
\includegraphics[width=0.70\textwidth,clip=true,angle=-90]{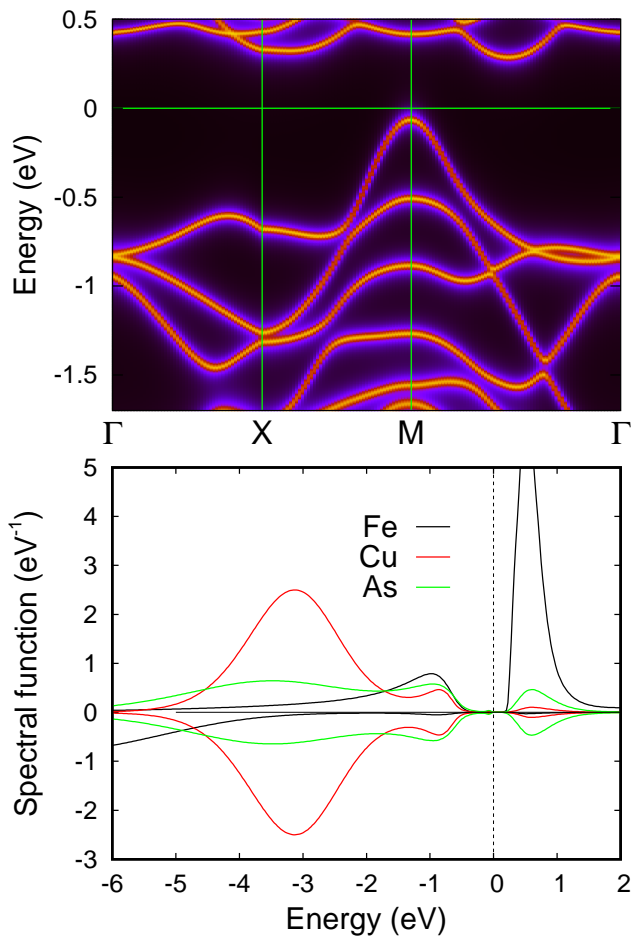}
\caption{
Electronic structure along the $\Gamma$-X-M-$\Gamma$ path (upper panel) and 
spin-resolved atomic-projected spectral functions of AFM Na(Fe$_{0.5}$Cu$_{0.5}$)As 
obtained by DFT+DMFT at $T=290$~K.}
\label{Fig_6}
\end{figure}

We perform spin-polarized DFT+DMFT calculations of the spectral properties of 
Na(Fe,Cu)As with Cu $x=0.5$. Our results for the orbitally-resolved Fe $3d$, Cu $3d$, 
and As $4p$ spectra are shown in Fig.~\ref{Fig_6} along with the {\bf k}-resolved 
spectral functions of AFM Na(Fe$_{0.5}$Cu$_{0.5}$)As obtained by DFT+DMFT. Our 
results exhibit Mott-Hubbard insulating behavior with a band  gap of $\sim$0.3~eV, 
in agreement with experiments and previous theoretical estimates \cite{Matt_PRL_2016,Song_ncomm_2016}. 
Interestingly, Na(Fe,Cu)As with long-range antiferromagnetic ordering reveals a 
more coherent electronic structure near the Fermi level as compared to that in the 
PM state, consistent with previous studies \cite{Charnukha_PRB_2017,Mandal_PRB_2019}. 
Our result for the local magnetic moment of about 3.65~$\mu_{\rm B}$ (fluctuating 
moment 3.46~$\mu_{\rm B}$) is compatible with that in our model DMFT calculations 
for Cu doping $x=0.5$. The spectral function of Na(Fe,Cu)As with Cu $x=0.5$ shows 
a pronounced $\sim5$-6~eV splitting of the Fe $3d$ spin up and down states due to 
the magnetic exchange. Moreover, the spin-polarized DFT+DMFT calculations give a 
large ordered magnetic moment of 3.61 $\mu_{\rm B}$ per Fe site. We note that this 
value of the ordered magnetic moment is significantly larger than that reported 
from neutron scattering, 1.1~$\mu_{\rm B}$/Fe, suggesting the crucial role of the 
nonlocal correlation effects and Cu/Fe disorder in Na(Fe,Cu)As~\cite{Xin_PRB_2019}. 
Our results highlight the key role of electronic correlations while antiferromagnetic 
order alone within DFT could not open a gap in the electronic structure of 
NaFe$_{0.5}$Cu$_{0.5}$As. In addition, in DFT we observe that the overall bandwidth 
of heavily Cu doped NaFeAs is decreased from that in stoichiometric NaFeAs by $\sim 20$\%, 
triggering a Mott transition, in agreement with previous estimates \cite{Song_ncomm_2016}.

By taking into account both the local potential difference between the Cu and Fe 
ions and on-site Coulomb correlations we obtain a much better agreement between the  
{\bf k}-resolved spectral function of Na(Fe,Cu)As with $x=0.5$ and ARPES~\cite{Matt_PRL_2016} 
compared to the model Hamiltonian results in Sec.~\ref{sec_tb}. In particular, 
the spectral weight at the $\Gamma$ point forms a weakly dispersive band in the 
energy interval from -0.5 $-$ -0.7 eV along the $\Gamma$-X direction, in agreement 
with ARPES measurements \cite{Matt_PRL_2016}. We also observe a dispersive convex 
band along the X-M-$\Gamma$ line at the top of the valence band, which was absent 
in the model DMFT result for PM Na(Fe,Cu)As. It is also interesting to note high 
coherence of the electronic states of AFM NaFe$_{0.5}$Cu$_{0.5}$As to that in the 
PM phase (see Fig.~\ref{Fig_2}). 

We observe a remarkable sensitivity of the electronic structure and magnetic correlations 
in NaFeAs with respect to doping with Cu. In Fig.~\ref{Fig_5} we display the in-plain 
magnetic states and exchange couplings of AFM NaFe$_{1-x}$Cu$_x$As for Cu $x=0$ and 0.5. 
Our results for magnetic exchange couplings obtained from the spin-polarized DFT+DMFT 
magnetic force theorem calculations (for the Wannier Fe $3d$ states) show a relatively large 
antiferromagnetic inter-site exchange coupling $J_{1a}\sim 23$~meV (assuming an effective 
$S=2$ state per Fe ion) in AFM NaFe$_{0.5}$Cu$_{0.5}$As at a temperature 290~K~\cite{leip,H_Heisenberg}. 
Our results for the exchange couplings in AFM NaFeAs ($x=0$) are $J_{1a} = 31$~K and 
$J_{1b} = -22$~K for the nearest-neighbour and $J_2 = 11$~K for the next-nearest-neighbor 
(here we assume a $S=1/2$ state) as obtained by the spin-polarized DFT+DMFT at $T=145$ K. 
This result is compatible with that obtained by a more precise spin-wave calculations~\cite{Zhang_PRL_2014, Han_PRL_2009}. 
The calculated local (fluctuating) and ordered magnetic moments in AFM NaFeAs are 2.03 
and 0.77$\mu_\mathrm{B}$/Fe, respectively. Moreover, in stoichiometric NaFeAs the DFT+DMFT 
magnetization is found to sharply collapse to the PM state at temperatures above 145~K. 
Our results therefore suggest an intermediate range of correlation strength pointing out 
to itinerant nature of magnetic moments in pure NaFeAs.   

\section{Conclusion}
In conclusion, using DFT+DMFT we explored the effects of Coulomb correlations and 
hole doping on the electronic structure and magnetic properties of Cu-doped NaFeAs. 
Upon Cu-doping, we observe a strong orbital-dependent localization of the Fe $3d$ 
states accompanied by a large renormalization of electronic mass of the Fe $xy$ and 
$xz$/$yz$ states. Na(Fe,Cu)As shows bad metal behavior associated with a coherence-to-incoherence 
crossover of the Fe $3d$ electronic states and local moments formation. For Cu $x>0.375$ 
it is found to undergo a Mott-Hubbard metal-insulator transition which is accompanied 
by divergence of the quasiparticle effective mass of the Fe $xy$ states. In contrast 
to this, the $xz$/$yz$ and $e_g$ quasiparticle weights remain finite at the MIT. 
The Mott transition occurs via an intermediate orbital-selective Mott phase to appear 
at Cu $x\simeq 0.375$ characterized by Mott localized Fe $xy$ orbitals. Our DFT+DMFT 
results suggest a crucial importance of electron-electron correlations and local potential 
difference between the Cu and Fe ions in Na(Fe,Cu)As. We propose a possible importance 
of Fe/Cu disorder to explain the magnetic properties of Cu-doped NaFeAs.


\begin{thebibliography}{99}

\bibitem{review_cusc}
E. Dagotto, Rev. Mod. Phys. \textbf{66}, 763 (1994);
A. Damascelli, Z. Hussain, and Z.-X. Shen, Rev. Mod. Phys. \textbf{75} 473 (2003);
D. N. Basov and T. Timusk, Rev. Mod. Phys. \textbf{77}, 721 (2005);
P. A. Lee, N. Nagaosa, and X.-G. Wen, Rev. Mod. Phys. \textbf{78}, 17 (2006);
N. P. Armitage, P. Fournier, and R. L. Greene, Rev. Mod. Phys. {\bf 82}, 2421 (2010);
B. Keimer, S. A. Kivelson, M. R. Norman, S. Uchida, and J. Zaanen 
Nature \textbf{518}, 179 (2015).

\bibitem{review_fesc}
M. V. Sadovskii, Phys. Usp. {\bf 51} 1201 (2008);
I. I. Mazin, 
Nature \textbf{464}, 183 (2010);
J. Paglione and R. L. Greene, 
Nat. Phys. \textbf{6}, 645 (2010);
G. R. Stewart, 
Rev. Mod. Phys. {\bf 83}, 1589 (2011);
D. N. Basov and A. V. Chubukov, 
Nat. Phys. {\bf 7}, 272 (2011);
P. Dai, J. Hu, and E. Dagotto, 
Nat. Phys. {\bf 8}, 709 (2012);
P. C. Dai, 
Rev. Mod. Phys. \textbf{87}, 855 (2015);
Q. Si, R. Yu, and E. Abrahams, 
Nat. Rev. Mater. \textbf{1}, 16017 (2016);
R. M. Fernandes and A. V. Chubukov, 
Rep. Prog. Phys. \textbf{80}, 014503 (2017).

\bibitem{nematicity}
R. M. Fernandes, A. V. Chubukov, and J. Schmalian, 
Nat. Phys. {\bf 10}, 97 (2014).
J. K. Glasbrenner, I. I. Mazin, Harald O. Jeschke, P. J. Hirschfeld, 
R. M. Fernandes and R. Valent\'i, 
Nat. Phys. {\bf 11}, 953 (2015).

\bibitem{s_pm}
I. I. Mazin, D. J. Singh, M. D. Johannes, and M. H. Du, 
Phys. Rev.Lett. {\bf 101}, 057003 (2008);
A. V. Chubukov, D. V. Efremov, and I. Eremin, 
Phys. Rev. B {\bf 78}, 134512 (2008);
P. J. Hirschfeld, M. M. Korshunov, and I. I. Mazin, 
Rep. Prog. Phys. \textbf{74}, 124508 (2011);
A. V. Chubukov, Annu. Rev. Condens. 
Matter Phys. \textbf{3}, 57 (2012).

\bibitem{s_pm_2}
P. J. Hirschfeld, D. Altenfeld, I. Eremin, and I. I. Mazin, 
Phys. Rev. B {\bf 92}, 184513 (2015); 
D. Altenfeld, P. J. Hirschfeld, I. I. Mazin, and I. Eremin, 
Phys. Rev. B {\bf 97}, 054519 (2018).

\bibitem{Sprau_Science_2017}
P. O. Sprau, A. Kostin, A. Kreisel, A. E. B\"ohmer, V. Taufour, P. C. Canfield, 
S. Mukherjee, P. J. Hirschfeld, B. M. Andersen, J. C. S\'eamus Davis, 
Science \textbf{357}, 75 (2017);
A. Kostin, P. O. Sprau, A. Kreisel, Y. X. Chong, A. E. B\"ohmer, 
P. C. Canfield, P. J. Hirschfeld, B. M. Andersen, and J. C. S\'eamus Davis, 
Nat. Mater. \textbf{17}, 869 (2018).

\bibitem{pairing_cusc}
C. C. Tsuei and J. R. Kirtley, 
Rev. Mod. Phys. {\bf 72}, 969 (2000).

\bibitem{FeSCS_correlations}
Q. Si and E. Abrahams, 
Phys. Rev. Lett. \textbf{101}, 076401 (2008);
M. M. Qazilbash, J. J. Hamlin, R. E. Baumbach, L. Zhang, D. J. Singh, 
M. B. Maple, and D. N. Basov, 
Nat. Phys. \textbf{5}, 647 (2009).

\bibitem{Weber_NatPhys_2010}
C. Weber, K. Haule and G. Kotliar, 
Nat. Phys. {\bf 6}, 574 (2010).

\bibitem{PhysRevLett.110.146402}
I. Misawa, K. Nakamura, and  M. Imada, 
Phys. Rev. Lett. \textbf{108}, 177007 (2012);
R. Yu and Q. Si, 
Phys. Rev. Lett. \textbf{110}, 146402 (2013).

\bibitem{Deng_PRL_2013}
X. Deng, J. Mravlje, R. Zitko, M. Ferrero, G. Kotliar, and A. Georges, 
Phys. Rev. Lett. {\bf 110}, 086401 (2013).

\bibitem{Medici_PRL_2014}
L. de Medici, G. Giovannetti, and M. Capone, 
Phys. Rev. Lett. {\bf 112}, 177001 (2014).

\bibitem{Yu_PRB_2011}
R. Yu and Q. Si, 
Phys. Rev. B {\bf 84}, 235115 (2011).

\bibitem{dftdmft_fech}
S. Mandal, R. E. Cohen, and K. Haule
Phys. Rev. B {\bf89}, 220502(R) (2014);
%
I. Leonov, S. L. Skornyakov, V. I. Anisimov, and D. Vollhardt, 
Phys. Rev. Lett. {\bf 115}, 106402 (2015);
%
S. L. Skornyakov, V. I. Anisimov, D. Vollhardt, and I. Leonov, 
Phys. Rev. B {\bf 96}, 035137 (2017);
%
M. D. Watson, S. Backes, A. A. Haghighirad, M. Hoesch, T. K. Kim, A. I. Coldea, and Roser Valent\'i, 
Phys. Rev. B {\bf 95}, 081106(R) (2017);
%
S. L. Skornyakov, V. I. Anisimov, D. Vollhardt, and I. Leonov, 
Phys. Rev. B {\bf 97}, 115165 (2018);
%
S. L. Skornyakov and I. Leonov, 
Phys. Rev. B {\bf 100}, 235123 (2019).

\bibitem{Imada_review}
M. Imada, A. Fujimori, and Y. Tokura
Rev. Mod. Phys. \textbf{70}, 1039 (1998).

\bibitem{Nica_npj_2017}
E. M. Nica, R. Yu, and Q. Si, 
npj Quant. Mater. {\bf 2}, 24 (2017).

\bibitem{Song_ncomm_2016}
Y. Song, Z. Yamani, C. Cao, Y. Li, C. Zhang, J. S. Chen, Q. Huang, 
H. Wu, J. Tao, Y. Zhu, W. Tian, S. Chi, H. Cao, Y.-B. Huang, M. Dantz, 
T. Schmitt, R. Yu, A. H. Nevidomskyy, E. Morosan, Q. Si, and P. Dai,
Nat. Commun. {\bf 7}, 13879 (2016).

\bibitem{Wang_PRB_2013}
A. F. Wang, J. J. Lin, P. Cheng, G. J. Ye, F. Chen, J. Q. Ma, X. F. Lu, 
B. Lei, X. G. Luo, and X. H. Chen, 
Phys. Rev. B {\bf 88}, 094516 (2013).

\bibitem{Matt_PRL_2016}
C. E. Matt, N. Xu, B. Lv, J. Ma, F. Bisti, J. Park, T. Shang, C. Cao, 
Yu Song, A. H. Nevidomskyy, P. Dai, L. Patthey, N. C. Plumb, M. Radovic, 
J. Mesot, and M. Shi, 
Phys. Rev. Lett. {\bf 117} 097001 (2016).

\bibitem{Cun_PRX_2015}
C. Ye, W. Ruan, P. Cai, X. Li, A. Wang, X. Chen, and Y. Wang, 
Phys.Rev. X {\bf 5}, 021013 (2015).

\bibitem{Wright_PRB_2012}
J. D. Wright, T. Lancaster, I. Franke, A. J. Steele, J. S. M\"oller, 
M. J. Pitcher, A. J. Corkett, D. R. Parker, D. G. Free, F. L. Pratt, 
P. J. Baker, S. J. Clarke, and S. J. Blundell, 
Phys. Rev. B {\bf 85}, 054503 (2012).

\bibitem{Park_PRB_2012}
J. T. Park, G. Friemel, T. Loew, V. Hinkov, Yuan Li, B. H. Min, D. L. Sun, 
A. Ivanov, A. Piovano, C. T. Lin, B. Keimer, Y. S. Kwon, and D. S. Inosov
Phys. Rev. B {\bf 86}, 024437 (2012).

\bibitem{Li_PRB_2009}
S. Li, C. de la Cruz, Q. Huang, G. F. Chen, T.-L. Xia, J. L. Luo, 
N. L. Wang, and Pengcheng Dai, 
Phys. Rev. B {\bf 80}, 020504(R) (2009).

\bibitem{Zhang_PRL_2014}
C. Zhang, L. W. Harriger, Z. Yin, W. Lv, M. Wang, G. Tan, Y. Song, 
D. L. Abernathy, W. Tian, T. Egami, K. Haule, G. Kotliar, and P. Dai, 
Phys. Rev. Lett. {\bf 112}, 217202 (2014).

\bibitem{Tan_PRB_2017}
G. Tan, Y. Song, R. Zhang, L. Lin, Z. Xu, L. Tian, S. Chi, M. K. Graves-Brook, 
S. Li, and P. Dai,
Phys. Rev. B {\bf 95}, 054501 (2017).

\bibitem{DFTpU}
V. I. Anisimov, J. Zaanen, and O. K. Andersen
Phys. Rev. B {\bf 44}, 943 (1991);
V. I. Anisimov, F. Aryasetiawan, and A. I. Lichtenstein, 
J. Phys.: Condens. Matter {\bf 9}, 767 (1997).

\bibitem{Charnukha_PRB_2017}
A. Charnukha, Z. P. Yin, Y. Song, C. D. Cao, P. Dai, K. Haule, 
G. Kotliar, and D. N. Basov, 
Phys. Rev. B {\bf 96}, 195121 (2017).

\bibitem{dftdmft_nsc}
W. Metzner and D. Vollhardt, 
Phys. Rev. Lett. {\bf 62}, 324 (1989);
A. Georges, G. Kotliar, W. Krauth, and M. J. Rozenberg, 
Rev. Mod. Phys. {\bf 68}, 13 (1996);
%
V. I. Anisimov, A. I. Poteryaev, M. A. Korotin, A. O. Anokhin, and G. Kotliar,
J. Phys. Condens. Matter {\bf 9}, 7359 (1997);
%
G. Kotliar, S. Y. Savrasov, K. Haule, V. S. Oudovenko, O. Parcollet, and C. A. Marianetti,
Rev. Mod. Phys. {\bf 78}, 865 (2006);
%
A. I. Lichtenstein and M. I. Katsnelson, 
Phys. Rev. B {\bf 57}, 6884 (1998).
%
J. Kune\v s, I. Leonov, P. Augustinsk\' y, V. K\v r\' apek, M. Kollar, and D. Vollhardt, 
Eur. Phys. J. Spec. Top. {\bf 226}, 2641 (2017).

\bibitem{Xin_PRB_2020}
Y. Xin, I. Stolt, Y. Song, P. Dai, and W. P. Halperin,
Phys. Rev. B {\bf 101}, 064410 (2020).

\bibitem{U_in_superconductors}
K. Haule, J. H. Shim, and G. Kotliar,
Phys. Rev. Lett. {\bf 100}, 226402 (2008);
%
M. Aichhorn, L. Pourovskii, V. Vildosola, M. Ferrero,
O. Parcollet, T. Miyake, A. Georges, and S. Biermann,
Phys. Rev. B {\bf 80}, 085101 (2009);
%
S. L. Skornyakov, A. A. Katanin, and V. I. Anisimov,
Phys. Rev. Lett. {\bf 106}, 047007 (2011);
%
S. L. Skornyakov, V. I. Anisimov, and D. Vollhardt, 
Phys. Rev. B {\bf 86}, 125124 (2012);
%
M. Aichhorn, L. Pourovskii, and A. Georges,
Phys. Rev. B {\bf 84}, 054529 (2011);
%
J. M. Tomczak, M. van Schilfgaarde, and G. Kotliar,
Phys. Rev. Lett. {\bf 109}, 237010 (2012);
%
Z. P. Yin, K. Haule, and G. Kotliar,
Phys. Rev. B {\bf 86}, 195141 (2012);
%
P. Werner, M. Casula, T. Miyake, F. Aryasetiawan, A. J. Millis,
and S. Biermann,
Nat. Phys. {\bf 8}, 331 (2012);
%
A. Georges, L. de' Medici, and J. Mravlje,
Annu. Rev. Condens. Matter Phys. {\bf 4}, 137 (2013);
%
M. Hirayama, T. Miyake, and M. Imada,
Phys. Rev. B {\bf 87}, 195144 (2013);
%
I. A. Nekrasov, N. S. Pavlov, and M. V. Sadovskii, 
Jetp Lett. {\bf 102}, 26 (2015);
%
A. van Roekeghem, L. Vaugier, H. Jiang, and S. Biermann,
Phys. Rev. B \textbf{94}, 125147 (2016).

\bibitem{dmft_fesc}
Z. P. Yin, K. Haule, and G. Kotliar,
Nat. Phys. {\bf 7}, 294 (2011);
%
Z. P. Yin, K. Haule, and G. Kotliar,
Nat. Mater. {\bf 10}, 932 (2011).

\bibitem{WannierH}
N. Marzari, A. A. Mostofi, J. R. Yates, I. Souza, and D. Vanderbilt, 
Rev. Mod. Phys. \textbf{84}, 1419 (2012);
V. I. Anisimov, D. E. Kondakov, A. V. Kozhevnikov, I. A. Nekrasov \emph{et al.},
Phys. Rev. B {\bf 71}, 125119 (2005);
%
Dm. Korotin, A. V. Kozhevnikov, S. L. Skornyakov, I. Leonov, N. Binggeli,
V. I. Anisimov, and G. Trimarchi,
Eur. Phys. J. B {\bf 65}, 91 (2008);
%
G. Trimarchi, I. Leonov, N. Binggeli, Dm. Korotin, and V. I. Anisimov,
J. Phys.: Condens. Matter {\bf 20}, 135227 (2008).

\bibitem{Leonov_prb_2015}
I. Leonov, V. I. Anisimov, and D. Vollhardt,
Phys. Rev. B {\bf 91}, 195115 (2015).

\bibitem{dftdmft_sc}
I. Leonov, N. Binggeli, Dm. Korotin, V. I. Anisimov, N. Stoji\'c, and D. Vollhardt, 
Phys. Rev. Lett. \textbf{101}, 096405 (2008);
I. Leonov, Dm. Korotin, N. Binggeli, V. I. Anisimov, and D. Vollhardt, 
Phys. Rev. B \textbf{81}, 075109 (2010);
%
I. Leonov,
Phys. Rev. B \textbf{92}, 085142 (2015);
%
I. Leonov, L. Pourovskii, A. Georges, and I. A. Abrikosov,
Phys. Rev. B \textbf{94}, 155135 (2016);
I. Leonov, A. O. Shorikov, V. I. Anisimov, and I. A. Abrikosov,
Phys. Rev. B \textbf{101}, 245144 (2020).

\bibitem{GGA}
S. Baroni, S. de Gironcoli, A. Dal Corso, and P. Giannozzi,
Rev. Mod. Phys. {\bf 73}, 515 (2001);
%
P. Giannozzi, S. Baroni, N. Bonini, M. Calandra, R. Car {\it et al.},
J. Phys.:Condens. Matter {\bf 21}, 395502 (2009).

\bibitem{ctqmc}
P. Werner, A. Comanac, L. de Medici, M. Troyer, and A. J. Millis,
Phys. Rev. Lett. {\bf 97}, 076405 (2006);
%
E. Gull, A. J. Millis, A. I. Lichtenstein, A. N. Rubtsov,
M. Troyer, and P. Werner,
Rev. Mod. Phys. {\bf 83}, 349 (2011).

\bibitem{Pade}
H. J. Vidberg and J. W. Serene,
J. Low Temp. Phys. {\bf 29}, 179 (1977).

\bibitem{Yu_PRB_2017}
R. Yu and Q. Si, 
Phys. Rev. B {\bf 96}, 125110 (2017).

\bibitem{brinkman_rice}
W. F. Brinkman and T. M. Rice, 
Phys. Rev. B {\bf 2}, 4302 (1970).

\bibitem{fesc_pipi}
T. J. Liu, J. Hu, B. Qian, D. Fobes, Z. Q. Mao, W. Bao, M. Reehuis, 
S. A. J. Kimber, K. Proke\v s, S. Matas, D. N. Argyriou, A. Hiess, 
A. Rotaru, H. Pham, L. Spinu, Y. Qiu, V. Thampy, A. T. Savici, 
J. A. Rodriguez and C. Broholm,  
Nat. Mater. {\bf 9}, 718 (2010);
%
G. S. Tucker, D. K. Pratt, M. G. Kim, S. Ran, A. Thaler,
G. E. Granroth, K. Marty, W. Tian, J. L. Zarestky, M. D.
Lumsden, S. L. Bud’ko, P. C. Canfield, A. Kreyssig, A. I.
Goldman, and R. J. McQueeney, 
Phys. Rev. B {\bf 86}, 020503 (R) (2012);
%
Y. Texier, Y. Laplace, P. Mendels, J. T. Park, G. Friemel,
D. L. Sun, D. S. Inosov, C. T. Lin, and J. Bobroff,
Europhys. Lett. {\bf 99}, 17002 (2012);
%
R. M. Fernandes and A. J. Millis, 
Phys. Rev. Lett. {\bf 110}, 117004 (2013).

\bibitem{leip}
A. I. Liechtenstein, M. I. Katsnelson, V. P. Antropov, and
V. A. Gubanov, J. Magn. Magn. Mater. {\bf 67}, 65 (1987);
Y. O. Kvashnin, O. Gr\aa{}n\"as, I. Di Marco, M. I. Katsnelson,
A. I. Lichtenstein, and O. Eriksson, Phys. Rev. B {\bf 91}, 125133 (2015);
Dm. M. Korotin, V. V. Mazurenko, V. I. Anisimov, and S. V. Streltsov, 
Phys. Rev. B \textbf{91}, 224405 (2015).

\bibitem{Mandal_PRB_2019}
S. Mandal, K. Haule, K. M. Rabe, and D. Vanderbilt, 
Phys. Rev. B {\bf 100}, 245109 (2019).

\bibitem{Xin_PRB_2019}
Y. Xin, I. Stolt, J. A. Lee, Y. Song, P. Dai, and W. P. Halperin, 
Phys. Rev. B {\bf 99}, 155114 (2019). 

\bibitem{H_Heisenberg}
Here we use the following notation for the Heisenberg model $H=\sum_{i>j}J_{ij}{\bf S}_{i}{\bf S}_{j}$.

\bibitem{Han_PRL_2009}
M. J. Han, Q. Yin, W. E. Pickett, and S. Y. Savrasov,
Phys. Rev. Lett. {\bf 102}, 107003 (2009).

\end{thebibliography}
\end{document}